\documentclass[article,oneside]{article}

\usepackage{include/cs-seminar}
\usepackage{include/lstcoq}
\usepackage{include/lstisabelle}

\usepackage{kpfonts}
\usepackage{inconsolata}

\usepackage[medium]{titlesec}
\usepackage{listings}  % already imported in cs-seminar.sty, here is needed for correct syntax highlighting in TeXstudio.

\usepackage{enumitem}  % remove spaces before enumerations and lists
\setlist{leftmargin=4mm}
\newlist{subitemize}{itemize}{3}
\setlist[subitemize]{label=$\circ$,nolistsep}

\usepackage[compatibility=false]{caption}

\usepackage[sc]{mathpazo} % Use the Palatino font
\usepackage[T1]{fontenc} % Use 8-bit encoding that has 256 glyphs
\usepackage{microtype} % Slightly tweak font spacing for aesthetics

\usepackage[english]{babel} % Language hyphenation and typographical rules

\usepackage[hmarginratio=1:1,top=32mm,columnsep=20pt]{geometry} % Document margins
\usepackage{caption} % Custom captions under/above floats in tables or figures
\usepackage{booktabs} % Horizontal rules in tables

\usepackage{enumitem} % Customized lists
\setlist[itemize]{noitemsep} % Make itemize lists more compact

\usepackage{abstract} % Allows abstract customization
\usepackage{fancyhdr} % Headers and footers
\usepackage{hyperref}

\usepackage{authblk}

\title{\bf Comparison of Two Theorem Provers:\protect\\ Isabelle/HOL and Coq}
\author[1,2]{Artem Yushkovskiy}
\affil[1]{Department of Computer Science,\protect\\ Aalto University (Espoo, Finland)
\vspace{1em}}
\affil[2]{Department of Computer System Design and Security,\protect\\ ITMO University (St. Petersburg, Russia)}
\date{\vspace{1em}Autumn 2017}

\begin{document}
%\counterwithin{lstlisting}{section}

\maketitle

\begin{abstract}

The need for formal definition of the very basis of mathematics arose in the last century.
The scale and complexity of mathematics, along with discovered paradoxes, revealed the danger of accumulating errors across theories. Although, according to G\"odel's incompleteness theorems, it is not possible to construct a single formal system which will describe all phenomena in the world, being complete and consistent at the same time, it gave rise to rather practical areas of logic, such as the theory of automated theorem proving. This is a set of techniques used to verify mathematical statements mechanically using logical reasoning. Moreover, it can be used to solve complex engineering problems as well, for instance, to prove the security properties of a software system or an algorithm.
This paper compares two widespread tools for automated theorem proving, Isabelle/HOL~\cite{tool_Isabelle} and Coq~\cite{tool_Coq}, with respect to expressiveness, limitations and usability. For this reason, it firstly gives a brief introduction to the bases of formal systems and automated deduction theory, their main problems and challenges, and then provides detailed comparison of most notable features of the selected theorem provers with support of illustrative proof examples.

\vspace{3mm}
\noindent KEYWORDS: proof assistants, Coq, Isabelle/HOL, logics, proof theory, formal method, classical logic, intuitionistic logic, usability.
\end{abstract}

%============================================================

\section{Introduction}
\label{sec:introduction}

Nowadays, the search for foundations of mathematics has become one of the key questions in philosophy of mathematics, which eventually has an impact on numerous problems in modern life. As a result, the \textit{formal approach} was developed as a new methodology for manipulating the abstract essences in a verifiable way. In other words, it is possible to follow the sequence of such manipulations in order to check the validity of each statement and, as a result, of a system at whole. Moreover, automating such a verification process can significantly increase reliability of formal models and systems based on them.

At present, a large number of tools have been developed to automate this process. Generally, these tools can be divided into two broad classes. 
The first class contains tools pursuing the aim of validating the input statement (\textit{theorem}) with respect to the sequence of inference transitions (user-defined \textit{proof}) according to set of inference rules (defined by logic). Such tools are sometimes called \textit{proof assistants}, their purpose is to help users to develop new proofs. The tools \textit{Isabelle}~\cite{tool_Isabelle}, \textit{Coq}~\cite{tool_Coq}, \textit{PVS}~\cite{tool_Pvs} are well-known examples of such systems, which are commonly used in recent years.

The second class consists of tools that automatically \textit{discover} the formal proof, which can rely either on induction, on meta argument, or on higher-order logic. Such tools are often called \textit{automated theorem provers}, they apply techniques of automated logical reasoning to develop the proof automatically. The systems \textit{Otter}~\cite{tool_Otter} and \textit{ACL2}~\cite{tool_Acl} are commonly known examples of such tools.
In this paper, only systems of the first class were considered in order to test the usability of such systems.

This paper is organised as follows. Section~\ref{sec:formal_theory} describes basic foundations of logic necessary for understanding theorem provers. In particular, Section~\ref{sec:definitions} provides formal definition, Sections~\ref{sec:properties}--\ref{sec:type_systems} describe different types, basic properties and theoretical limitations of formal systems.
Section~\ref{sec:comparison} presents the comparison itself and provides the illustrative examples of different kinds of proofs in both considering systems.

%------------------------------------------------------------

\subsection{Related work}
A considerably extensive survey on theorem provers has been presented by F.~Wiedijk~\cite{Wie03}, where fifteen most common systems for the formalization of mathematics were compared against various properties, in particular, size of supporting libraries, expressiveness of underlying logic, size of proofs (the de Bruijn criterion) and level of automation (the Poincaré principle). 
Another notable work was presented by D.~Griffioen and M.~Huisman~\cite{Griff98}, in which two theorem provers, PVS and Isabelle/HOL, were deeply compared with respect to numerous important aspects, such as properties of used logic, specification language, user interface, etc. This paper proposes analogous comparison of two widely used theorem provers, Isabelle and Coq, with respect to expressiveness, limitations and usability.

%============================================================

\section{Foundations of formal approach}
\label{sec:formal_theory}

The formal approach appeared in the beginning of previous century when mathematics experienced deep fundamental crisis caused by the need for a formal definition of the very basis. At that time, multiple paradoxes in several fields of mathematics have been discovered. Moreover, the radically new theories appeared just by modification of the set of axioms, e.g., reducing the parallel postulate of Euclidean geometry has lead to completely different non-Euclidian geometries, such as Lobachevsky's hyperbolic geometry or Riemman's elliptic geometry, that eventually have a large number of applications in both natural sciences and engineering.

\subsection{Definition of the formal system}
\label{sec:definitions}

% TODO: add citations!

Let the \textit{judgement} be an arbitrary statement. The \textit{formal proof} of the formula $\phi$ is a finite sequence of judgements $ ( \psi_i )_{i=1}^{n} $, where each $\psi_i$ is either an axiom $A_i$, or a formula inferred from the subset $\{ \psi_k \}_{k=1}^{i-1}$ of previously derived formulas according the \textit{rules of inference}. \textit{An axiom} $A_i \in A$ is a judgement evidently claimed to be true. \textit{A logical inference} is a transfer from one judgement (\textit{premise}) to another (\textit{consequence}), which preserves truth. In formal logic, inference is based entirely on the structure of those judgements, thereby, the result formal system represents the abstract model describing part of real world.

The formulas consist of \textit{propositional variables}, connected with \textit{logical connectives} (or logical operators) according to rules, defined by a formal language. The formulas, which satisfy such rules, are called \textit{well-formed formulas} (wff). Only wff can form judgements in a formal system. The propositional variable is an atomic formula that can be claimed as either true or false. The logical connective is a symbol in formal language that transforms one wff to another. Typically, the set of logical connectives contains negation $\neg$, conjunction $\land$, disjunction $\lor$, and implication $\rightarrow$ operators, although the combination of negation operator with any other of aforementioned operators will be already functionally complete (i.e., any formula can be represented with the usage of these two logical connectives).

The formal system described above does not contain any restriction on the form of propositional variables, such logic is called \textit{propositional logic}. However, if these variables are quantified on the sets, such logic is called \textit{first-order} or \textit{predicate logic}. Commonly, first-order logic operates with two quantifiers, the universal quantifier $\forall$ and the existential quantifier~$\exists$. Thereafter, the \textit{second-order logic} extends it by adding quantifiers over first-order quantified sets --- relations defining the sets of sets. In turn, it can be extended by the \textit{higher-order logic}, which contain quantifiers over the arbitrary nested sets (for instance, the expression $\forall f: bool \rightarrow bool, f\ (f\ (f\ x)) = f\ x$ could be considered in higher-order logic), or the \textit{type theory}, which assigns a type for every expression in the formal language (see Section~\ref{sec:type_systems}).

A formal system determines the set of derivable \textit{formulas} (judgements that are provable with respect to the rules of formal system). Let $\Phi$ be a set of formulas. Initially, it only consists of \textit{hypotheses}, a priori true formulas, which are claimed to be already proved. The notation $\Phi \vdash \phi$ means that the formula $\phi$ is \textit{provable} from $\Phi$, if there exists a proof that infers $\phi$ from $\Phi$. The formula which is provable without additional premises is called \textit{tautology} and denoted as $ \vdash \phi $ (meaning $\emptyset \vdash \phi$). The formula is called \textit{contradiction} if \ $\vdash \neg \phi$. Obviously, all contradictions are equivalent in one formal system, they are denoted as $\bot$.
%TODO: model, |= ...

In current paper, the notation~\eqref{notation_infrule}, which was borrowed from the Isabelle documentation, will be used for expressing the rules of inference. In this notation, the sign $\implies$ means logical implication, which is right-associative, see formula~\eqref{notation_impl_associative}. This notation is equivalent to the standard notation~\eqref{notation_standard}:
\begin{gather}
[\![ A_{1}; A_{2}; \dots A_{n} ]\!] \implies B 
    \label{notation_infrule}\\
\equiv \{ A_{1}, A_{2}, \dots A_{n} \} \vdash B
    \label{notation_standard}
\end{gather}
\begin{equation}\label{notation_impl_associative}
\begin{aligned}
A_{1} \implies A_{2} \implies \dots \implies A_{n} \implies B \\
\equiv  A_{1} \implies ( A_{2} \implies ( \dots \implies ( A_{n} \implies B)))
\end{aligned}
\end{equation}

The formulas below describe the principal inference rule residing in most logic systems, the \textit{Modus ponens} (MP) rule, and two main axioms of classical logic:
\begin{gather}
[\![ A, A \implies B ]\!] \implies B
    \label{rule_modus_ponens}\tag{MP} \\
A \implies (B \implies A).
	\label{axiom_hilbert_1}\tag{A1} \\
(A \implies (B \implies C)) \implies ((A \implies B) \implies (A \implies C)).
	\label{axiom_hilbert_2}\tag{A2}
\end{gather}

Together with axioms \eqref{axiom_hilbert_1} and \eqref{axiom_hilbert_2}, Modus ponens rule forms the Hilbert proof system which can process statements of classical propositional logic.
%is not complete under classical semantics.
Other classical logic systems often include the axiom of excluded middle~\eqref{axiom_excluded_middle}, and may derive the double negation introduction~\eqref{rule_double_negation_intr} and double negation elimination~\eqref{rule_double_negation_elim} laws:
\begin{gather}
A \lor \neg A.
	\label{axiom_excluded_middle}\tag{EM} \\
A \implies \neg \neg A
\label{rule_double_negation_intr}\tag{DNi} \\
\neg \neg A \implies A
\label{rule_double_negation_elim}\tag{DNe}
\end{gather}

Many classical logics may derive the de Morgan's laws~\eqref{tauto_demorgan1},~\eqref{tauto_demorgan2}, the law of contraposition~\eqref{tauto_contrapos}, the Peirce's law~\eqref{tauto_peirce} and many other tautologies:
\begin{gather}
\neg (A \land B) \Longleftrightarrow \neg A \lor \neg B 
    \label{tauto_demorgan1}\tag{DM1} \\
\neg (A \lor B) \Longleftrightarrow \neg A \land \neg B 
    \label{tauto_demorgan2}\tag{DM2} \\
(A \rightarrow B) \implies (\neg B \rightarrow \neg A) 
    \label{tauto_contrapos}\tag{CP} \\
((A \rightarrow B) \rightarrow A) \implies B
    \label{tauto_peirce}\tag{PL}
\end{gather}

The axiom of excluded middle means that every logical statement is decidable, which might not be true in some applications. Adding this axiom to the formal system leads to the reasoning from \textit{truth} statements, in contrast to \textit{natural deduction systems} that use reasoning from \textit{assumptions}. Although the difference between these two kinds of formal systems seems to be subtle, the latter can be used more as framework, allowing to build new systems on the logical base of pre-defined premises and formal proof rules. % AUTHOR'S THOUGHTS, UNVERIFIED

\subsection{Properties of Formal System}
\label{sec:properties}

Let $U$ be a set of all possible formulas, let $\Gamma = \ <A, V, \Omega, R>$ be a formal system with set of axioms $A$, set of propositional variables $V$, set of logical operators $\Omega$, and set of inference rules $R$. Then $\Gamma$ is called:
\begin{itemize}
	\itemsep0em
	\item \textit{consistent}, if both formula and its negation can not be proved in the system: \\
		$\nexists \phi \in \Gamma: \ \Gamma \vdash \phi \land \Gamma \vdash \neg \phi  \ \Leftrightarrow \ \Gamma \nvdash \bot$;
	\item \textit{complete}, if all true statements can be inferred: \\
		$\forall \phi \in U: \ A \vdash \phi \lor A \vdash \neg \phi$ ;
	\item \textit{independent}, if no axiom can be inferred from another: \\
		$\not \exists a \in A: \ A \vdash a$.
\end{itemize}

For instance, the Hilbert system described above is consistent and independent, yet incomplete under the classical semantics. In 1931, Kurt Gödel proved his first incompleteness theorem which states that any consistent formal system is incomplete. Later, in 1936, Alfred Tarski extended this result by proving his Undefinability theorem, which states that the concept of truth cannot be defined in a formal system. 
% originally: in arithmetic. Does that make sense?
In that case, modern tools, such as Coq, often restrict propositions to be either provable or unprovable, rather than true or false.
%Note, that the first-order logic is \textit{undecidable}, so that there does not exist a decision algorithm which is sound, complete and terminating.

\subsection{Lambda-calculus}
\label{sec:lambda}

The necessity of building the automatic reasoning systems has lead to development of models that abstract the computation process. That time, the concept of effective computability was being evolving rapidly, causing development of multiple formalisations of computation, such as Turing Machine, Normal Markov algorithms, Recursive functions, and other. One of the fist and most effective models was $\lambda$-\textit{calculus} invented by Alonzo Church in 1930s. This formalism provides solid theoretical foundation for the family of functional programming languages~\cite{Roj15}. In $\lambda$-calculus, functions are first-order objects, which means functions can be applied as arguments to other functions.

The central concept in $\lambda$-calculus is an \textit{expression}, which can be defined as a subject for application the rewriting rules~\cite{Bar88}. The basic rewriting rules of $\lambda$-calculus are listed below:

\begin{itemize}
\itemsep0em
	\item \textit{application}:
	$f a$ is the call of function $f$ with argument $a$
	
	\item \textit{abstraction}:
	$\lambda x.t[x]$ is the function with formal parameter x and body $t[x]$
	
	\item \textit{computation} ($\beta$-\textit{reduction}): replace formal parameter $x$ with actual argument $a$: \\
	$(\lambda x.t[x]) a \rightarrow_{\beta} t[x:=a]$
\end{itemize}

$\lambda$-calculus described above is called the \textit{type-free} $\lambda$-calculus. The more strong calculi can be constructed by using the types of expressions to the system, for which some useful properties can be proven (e.g., termination and memory safety)~\cite{Bar13}.

\subsection{Type systems}
\label{sec:type_systems}

A \textit{type} is a collection of elements. In a type system, each element is associated with a type, which defines a basic structure of it and restricts set of possible operations with the element. This allows to reveal useful properties of the formal system. Therefore, type theory serves as an alternative to the classical set theory~\cite{Tho91}.
% remove notation below if not used:
%In our notation, the $a =_{\tau} b$ means that $a$ equals $b$ and both of them are of type $\tau$. This notation is used for convenience to encode information about type to equality.

The function that builds a new type from another is called \textit{type constructor}. Such functions have been used long before type theories had been constructed formally, even in the~19th~century Giuseppe Peano used type constructor $S$ called the \textit{successor} function, along with zero element 0, to axiomatise natural number arithmetic. Thus, number $3$ can be constructed as $S(S(S(0))))$.

\subsubsection{Simple type theory}
%In the context of current paper, it is useful to describe 
The type can be defined declaratively, by assigning a label to set of values. Such types are called \textit{simple types}, they can be useful to avoid some paradoxes of set theory, e.g., separating sets of individuals and sets of sets allows to avoid famous Russel's paradox~\cite{Irv95}. Simple type theory can extend $\lambda$-calculus to a higher-order logic through connection between formulas and expressions of type Boolean~\cite{Paulson90}.

% TO-DO: // == Pure Type System ?
% https://www.cl.cam.ac.uk/teaching/1516/Types/lectures/lecture-9.pdf
% https://en.wikiversity.org/wiki/Foundations_of_Functional_Programming/Pure_type_systems
% http://www4.di.uminho.pt/~mjf/pub/SFV-CIC-2up.pd
% TO-DO: // == Polymorphic Type Theory ?
%\subsubsection{Polymorphic Type Theory}
% //not needed

% --
% \subsubsection{System F}
% second-order lambda calculus, is a typed lambda calculus that differs from the simply typed lambda calculus by the introduction of a mechanism of universal quantification over types
% //not needed

\subsubsection{Martin-Löf type theory}
The Martin-Löf type theory, also known as the \textit{Intuitionistic type theory}\footnote{In this paper, in terms \textit{intuitionistic type theory} and \textit{intuitionistic logic}, the word \textit{intuitionistic} is used as a synonym for \textit{constructive}.}, is based on the principles of constructive mathematics, that require explicit definition of the way of "constructing" an object in order to prove its existence. Therefore, an important place in intuitionistic type theory is held by the \textit{inductive types}, which were constructed recursively using a basic type (zero) and successor function which defines "next" element.

The Intuitionistic type theory also uses a wide class of \textit{dependent types}, whose definition depends on a value. For instance, the $n$-ary tuple is a dependent type that is defined by the value of~$n$. However, the type checking for such a system is an undecidable problem since determining of the equality of two arbitrary dependent types turns to be tantamount to a problem of inducing the equivalence of two non-trivial programs (which is undecidable in general case according to the Rice's theorem~\cite{Rice53}).

%"""A language with dependent types may include references to programs inside of types. For instance, the type of an array might include a program expression giving the size of the array, making it possible to verify absence of out-of-bounds accesses statically. Dependent types can go even further than this, effectively capturing any correctness property in a type. PVS’s dependent types are much more general, but they are squeezed inside the single mechanism of subset types, where a normal type is refined by attaching a predicate over its elements. Each member of the subset type is an element of the base type that satisfies the predicate."""

\subsubsection{Calculus of Constructions}
% TODO: figure out how this implies to PTS - Pure Type Systems

Another important constructive type theory is the Calculus of Constructions~(CoC) developed by Thierry~Coquand and Gérard~Huet in 1985~\cite{Coq85}. It represents a natural deduction system which incorporates dependent types, polymorphism and type constructors. The typed polymorphic functional language of CoC allow to define inductive definitions, although rather inefficiently~\cite{Paulin15}.

% see also the Coq documentation: https://coq.inria.fr/refman/cic.html#conv-rules

Whenever an inductive type is defined, the task of \textit{type-checking} becomes equivalent to the task of executing corresponding function in a programming language. Although in many programming languages type-checking algorithm is efficient, type-checking in CoC is \textit{undecidable} in general case. This problem is closely related to the \textit{Curry-Howard isomorphism}, a direct relationship between a program and an intuitionistic proof, in which a base type of the program is equivalent to a propositional variable in the proof, an empty type represents \textit{false} and a singletone type represents \textit{truth}, a functional type $T_{1}~\rightarrow~T_{2}$ corresponds to an implication, a product type $T_{1} * T_{2}$ and a sum type $T_{1} + T_{2}$ correspond to conjunction and disjunction, respectively~\cite{Pierce2002}. Thus the Calculus of Constructions can be considered as an extension of the Curry-Howard isomorphism.
An important feature of CoC type system is that it holds the \text{strong normalisation property}, which means that every sequence of inference eventually terminates with an irreducible normal form.
%This property does not admit the definitions of infinitely recursive structures and functions.
% TODO: check whether one can say that with respect to any formal system, not only to an abstract rewriting system

%Inference rules for the Calculus of Constructions:
%\begin{enumerate}
%    \item ${\displaystyle {{} \over {}\Gamma \vdash P:T}}$
%    \item ${\displaystyle {\Gamma \vdash A:K \over {\Gamma ,x:A\vdash x:A}}}$
%    \item ${\displaystyle {\Gamma ,x:A\vdash B:K\qquad \Gamma ,x:A\vdash N:B \over {\Gamma \vdash (\lambda x:A.N):(\forall x:A.B):K}}}$
%    \item ${\displaystyle {\Gamma \vdash M:(\forall x:A.B)\qquad \Gamma \vdash N:A \over {\Gamma \vdash MN:B[x:=N]}}}$
%    \item ${\displaystyle {\Gamma \vdash M:A\qquad A=_{\beta }B\qquad B:K \over {\Gamma \vdash M:B}}}$ % \over {\Gamma \vdash M : B}}
%\end{enumerate}
%
% perhaps: add explanations to the inference rules or delete them

Although the language of CoC is rather expressive, its expressiveness is not enough to prove some natural properties of types. In order to overcome this drawback, the \textit{Calculus of Inductive Constructions}~(CIC) was developed by Christine Paulin in 1990. CIC is implemented by adding the Martin-Löf's primitive inductive definitions to the CoC in order to perform the efficient computation of functions over inductive data types in higher-order logic~\cite{Paulin15}. This formalism lies behind the Coq proof assistant.

%============================================================

% perhaps: exclude this paragraph.

%\section{Methods for automated reasoning}
%\label{sec:auto_reasoning}

%techniques in common words (and in introduced previously notation), e.g.: 
%\begin{itemize}
%\itemsep0em
%	\item Clause rewriting
%		Simplification - The concept of (conditional) term rewriting is introduced and its realization as the proof method simp is explained. (from http://isabelle.in.tum.de/coursematerial/PSV2009-1/)
		
%	\item Resolution
%	\item Sequent Deduction
%	\item Natural Deduction
%	\item The Matrix Connection Method
%	\item Term Rewriting (+lambda calculus)
%	\item Mathematical Induction
%\end{itemize}

%============================================================

\section{Comparison of two theorem provers}
\label{sec:comparison}

In this work, two automated proof assistants, \textit{Isabelle/HOL}\footnote{Roughly speaking, Isabelle is a core for an automated theorem proving which supports multiple logical theories: Higher-Order Logic (HOL), first-order logic theories such as Zermelo-Fraenkel Set Theory (ZF), Classical Computational Logic (CCL), etc. In this paper, the Isabelle/HOL has been considered as the startpoint for exploring the power of this proof assistant.} and \textit{Coq} have been chosen for comparison as they both are widely used tools for theorem proving (according to the number of theorems that have already been formalised, see~\cite{Wiedijk100}). 

This section discusses some common and different features of these two theorem provers, providing illustrative examples of proofs performed in Coq and Isabelle. As a startpoint, the de Morgan's laws~\eqref{tauto_demorgan1} and~\eqref{tauto_demorgan2} in propositional and first-order logics have been chosen. Afterwards, the formula for sum of first $n$ natural numbers, defined inductively in both Isabelle and Coq, is being discussed. An example of extraction in Coq the verified code in Haskell and OCaml follows the proof of correctness of this formula.

%------------------------------------------------------------

\subsection{The Isabelle/HOL theorem prover}
\label{sec:prover_isabelle}

\textit{Isabelle} was developed as a successor of HOL theorem prover~\cite{tool_HOL} by Larry Paulson at the University of Cambridge and Tobias Nipkow at Technische Universität München. Isabelle was released for the first time in 1986 (two years after the Coq's first release). It was built in a modular manner, i.e., around a relatively small core, which can be extended by numerous basic theories that describe logic behind Isabelle. In particular, the theory of higher-order logic is implemented as Isabelle/HOL, and it is commonly used because of its expressivity and relative conciseness. 

Isabelle exploits classical logic, so even propositional type is declared as a set of two elements \texttt{true} and \texttt{false} (thus any $n$-ary logic can be easily formalised). In proofs, Isabelle combines several languages: \textit{HOL} as a functional programming language (which must be always in quotes), and \textit{Isar} as the language for describing procedures in order to manipulate the proof.

% TERMINATION - not include in the paper yet
% wiki: """Isabelle's main proof method is a higher-order version of resolution, based on higher-order unification."""
%\subsubsection{Algorithm termination}
%// some words on termination checks in Isabelle (unlike Coq)
%Termination: "The method lexicographic_order is the default method for termination proofs.
% https://isabelle.in.tum.de/doc/functions.pdf page 4 - !!! (relation method, lexicographic order, ...)
%// termination of computation. see: % https://www.joachim-breitner.de/blog/732-Isabelle_functions__Always_total%2C_sometimes_undefined
%for Isabelle: see slide 28 % https://www.it.uu.se/education/phd_studies/phd_courses/gc0910/isabelle/slide2.pdf
%- Primitive-recursive with primrec
%Terminating by construction
%- Well-founded recursion with fun
%Automatic termination proof
%- Well-founded recursion with function
%User-supplied termination proof

%------------------------------------------------------------

\subsection{The Coq theorem prover}
\label{sec:prover_coq}

\textit{Coq} is another widespread proof assistant system that has been developed at INRIA (Paris, France) since 1984. Coq is based on Calculus of Inductive Constructions, an implementation of intuitionistic logic which uses inductive and dependent types. Nonetheless, Coq's logic may be easily extended to classical logic by assuming the excluded middle axiom~\eqref{axiom_excluded_middle}. A key feature of Coq is a capability of extraction of the verified program (in OCaml, Haskell or Scheme) from the constructive proof of its formal specification~\cite{Letouzey08}. This facilitates using Coq as a tool for software verification.
% TODO!!! surprisingly, Isabelle can extract programs as well: "Isabelle/HOL allows to turn executable specifications directly into code in SML, OCaml, Haskell, and Scala." from https://www.cl.cam.ac.uk/research/hvg/Isabelle/overview.html

Being based on the constructive foundation, Coq has two basic meta-types, \texttt{Prop} as a type of propositions, and \texttt{Set} as a type of other types.
Unlikely Isabelle's type system, the \texttt{True} and \texttt{False} propositions are defined as of type of \texttt{Prop}, so that in order to be valid they need to be either assumed or proven (see Appendix~\ref{apx:type_definitions} Fig.\ref{ex_typedef_prop_coq}). Nonetheless, Coq's library has the \texttt{bool} definition, which is of type of \texttt{Set} in the manner of Isabelle's proposition (as simple as enumeration of two elements, tertium non datur; see Appendix~\ref{apx:type_definitions} Fig.\ref{ex_typedef_bool_coq}).

In proofs, Coq combines two languages: \textit{Gallina}, a purely functional programming language, and \textit{Ltac}, a procedural language for manipulating the proof process. A statement for proof and structures it relies on are written in Gallina, while the proof process itself is being controlled by the commands written in Ltac language.

% TERMINATION - not include in the paper yet
%\subsubsection{Algorithm termination}
%// "Syntactic restriction on recursive calls on term":
%// Coq doesn't allow to define recursive functions without decreasing argument => always terminates

%------------------------------------------------------------

\subsection{Common features}

In general, both Isabelle and Coq work in a similar way: given the definition of a statement, they can either verify already written proof, or assist user in developing such a proof in an interactive fashion, so that the invalid proofs are not accepted.
During the proof process, the systems save the proof state, a set of \textit{premises} and set of \textit{goals} (the statements to be proved). Therefore, the proof may represent the sequence of \textit{tactics} applied to the proof state. A tactic may be thought as an inference rule, it can use already proved statements, remove hypotheses or introduce variables. Some tactics work on very high level, they can automatically solve complex equations or prove complex statements, so that the proof assistant acquires features of an automated theorem provers described in Section~\ref{sec:introduction}.

Both systems have rather large libraries with considerable amount of already proven lemmas and theorems; in addition, they can be used as functional programming languages as they allow to construct new data types and recursive functions, they have pattern matching, type inference and other features inherent for functional languages.

Both tools are being actively developed: on the moment of writing this paper (autumn 2017), the latest versions were Coq 8.7.0 (stable) and Isabelle2017, both released in October 2017. Since their first release, both Isabelle and Coq have already been used to formalize enormous amount of mathematical theorems, including those which have very large or even controversial proof, such as Four colour theorem (2004), Lax-Milgram theorem (2017), and other important theorems~\cite{Wiedijk100}. Moreover, the theorem provers have been successfully used for testing and verifying of software programs, including the general-purpose operating system kernel~seL4~(2009)~\cite{Klein09}, the C~standard~(2015)~\cite{Krebbers15}, and others.

Both Isabelle and Coq have their own Integrated Development Environment (IDE) to work in (gtk-based CoqIDE and jEdit Prover IDE, respectively). In general, both native IDEs of these theorem provers provide the facility for interactive executing scripts step-by-step while preserving the state of proof (\textit{environment}), which for each step describes the set of premises along with already proved statements (\textit{context}) and the set of statements to be proven (\textit{goals}). However, Isabelle's native IDE allows to change the proof state arbitrarily, in contrast to the CoqIDE, which provides only the capability of switching the proof state to backward or forward linearly. Alternatively, both considering theorem provers have numerous of plugins for many popular IDEs, for instance, the Proof General~\cite{tool_PG} is a plugin for Emacs, which supports numerous proof assistants. During the work on this paper, only native IDEs of each proof assistant have been used in order to minimize the impact of third-party tools to the research.

Both systems accept proofs written in an imperative fashion (\textit{forward proof}), i.e., such proof represents a sequence of tactic calls, that implicitly change the proof state at each step, compounded by the control-flow operators called \textit{tacticals}, that combine tactics together, separate their results, repeat calls, etc. In addition, the syntax of Isar permits writing goals explicitly in the proof (\textit{backward proof}, see Appendix~\ref{apx:sum_inductive} Fig.~\ref{ex_nat_sum_isabelle} and Fig.\ref{ex_morgan_quant_isabelle}).

%<already in text, see paragraph above> Isar accepts more relaxed syntax in sense that its proof may look more like mathematical proof because of writing the goals during the proof (e.g., '\isabelleinline{assume \<not> P x then have \<exists>x. \<not> P x}') and using the connectors that can be read conveniently by human (e.g., '\isabelleinline{then}' abbreviates '\isabelleinline{from this}', '\isabelleinline{hence}' expands to '\isabelleinline{then have}', etc.). In contrast, proofs written in Coq look more like programs written in imperative programming language: the sequence of states need to be executed in order to check how the directives (application of tactics) change the state of proof. On the other hand, the brevity of Gallina language may let the experienced user, that knows the syntax and tactics well, spend less time for writing the proof in Coq rather than in Isabelle.

%------------------------------------------------------------

\subsection{Major differences}

The key differences between Isabelle and Coq lie in differences between logical theories they based on. While Isabelle/HOL exploits higher order logic along with non-dependent types, Coq is based on intuitionistic logic, which does not include the axiom of excluded middle~\eqref{axiom_excluded_middle} essential for classical logics.
Consequently, the double negation elimination rule~\eqref{rule_double_negation_elim} does not hold, however the double negation introduction law~\eqref{rule_double_negation_intr} can be easily proven (see Figures~\ref{ex_double_neg_elim_coq} and~\ref{ex_double_neg_intro_coq}). This follows from the fact that, if a proposition is known as truth, then double negation works as in classic logic, but if the proposition truthfulness is to be proven from its double negation, then there is nothing known about the proposition itself so far.

\begin{raggedleft}
\begin{tabular}{p{.45\linewidth} p{.45\linewidth}}
%\begin{figure}[!h]
%\begin{minipage}[t]{.48\textwidth}
\begin{lstlisting}[language=coq,
    caption={Proof failure of the \eqref{rule_double_negation_elim} rule in Coq},
    label=ex_double_neg_elim_coq]
Lemma DoubleNegElim_Coq : forall P: Prop,
    ~~P -> P.
Proof.
    try tauto.  (* fails *)
Abort.


[*$ $*]
\end{lstlisting} % empty lines after code and [*$ $*] are needed for caption alignment
&
%\end{minipage}
%\begin{minipage}[t]{.48\textwidth}
\begin{lstlisting}[language=coq,
    caption={Proof of the \eqref{rule_double_negation_intr} rule in Coq},
    label=ex_double_neg_intro_coq]
Lemma DoubleNegIntro_Coq : forall P: Prop,
    P -> ~~P.
Proof.
    (* automatic 'tauto' works here *)
    unfold not.
    intros P P_holds P_impl_false.
    apply P_impl_false. apply P_holds. 
Qed.
\end{lstlisting}
%\end{minipage}
%\end{figure}
\end{tabular}
\end{raggedleft}

In addition, the double-negated axiom of excluded middle can be proven as well solely in intuitionistic logic, see Appendix~\ref{apx:double_negated} Fig.\ref{ex_double_neg_ex_mid_coq}. This is a way for embedding the classical propositional logic into intuitionistic logic and known as \textit{Glivenko's double-negation translation}~\cite{Glivenko29}, which maps all classical tautologies to intuitionistic ones by double-negating them. Furthermore, there are other schemes of the translation for other classical logics, such as Gödel-Gentzen translation, Kuroda's translation, etc.~\cite{Kolmogorov25}.

Therefore, numerous of theorems, such as the classical logic tautology Peirce's law~\eqref{tauto_peirce}, can not be proved in intuitionistic logic, while being valid in classical logic, which makes the latter strictly weaker~\cite{Rush14} and incomplete (Coq's tactic for automatic reasoning of propositional statements \texttt{tauto} fails to prove this automatically).

In classical logic, some proofs remain valid, yet completely inapplicable. For instance, the following non-constructive proof of the statement "\textit{there exist algebraic irrational numbers $x$ and $y$ such that $x^y$ is rational}" may serve as a classic example of it.
The proof relies on the axiom of excluded middle~\cite{Harrison09}. Consider the number $\sqrt{2}^{\sqrt{2}}$; if it is rational, then consider $x = \sqrt{2}$ and $y = \sqrt{2}$, which both are irrational; if $\sqrt{2}^{\sqrt{2}}$ is irrational, then consider $x = \sqrt{2}^{\sqrt{2}}$ and $y = \sqrt{2}$, so that $x^{y}$ is rational, q.e.d. Although this proof is clear and concise, it reveals no information about whether the number $\sqrt{2}^{\sqrt{2}}$ is rational or irrational. More importantly, it gives no algorithm for finding other such numbers. Therefore, the main purpose of constructive proofs is to define such a solution schema for a problem, in addition to proving the claim.
Commonly, the proofs of existence\footnote{as well as proofs of non-universally valid statements "$\neg \forall$", which in classical logics are equivalent to existence proofs "$\exists$".} of an element are non-constructive as in order to prove such a statement it is enough to find single valid example.

%Although there are some conceptual differences related to different underlying logic frameworks, most basic datatypes (\texttt{bool}, \texttt{nat}) are defined in similar way in both Coq and Isabelle. For instance, Coq has two basic meta-datatypes, \texttt{Prop} as a type of propositions, and \texttt{Set} as a type of other types, in contrast to Isabelle, which defines propositions as a \texttt{Set} type as well. 

% In Coq, "Implications are functions" => theorems statements may be seen as functions %TODO

%------------------------------------------------------------

\subsubsection{Proofs in propositional logic}

As an example of proof statement in propositional logic, the de Morgan's law~\eqref{tauto_demorgan2} has been chosen. Although both proof assistants can operate with propositional statements, the proof in Isabelle relies on the classical logic by applying excluded middle~\eqref{axiom_excluded_middle} axiom (see "\texttt{apply (rule classical)}", ~Appendix~\ref{apx:morgan_propos} Fig.\ref{ex_morgan_propos_isabelle}), and the proof in Coq does not use this axiom, working completely within intuitionistic logic with propositional variables of meta-type \texttt{Prop} (see Appendix~\ref{apx:morgan_propos} Fig.\ref{ex_morgan_propos_coq}). Note that both system can prove this statement automatically (using tactic \texttt{blast} in Isabelle or tactic \texttt{tauto} in Coq).
 
%Both systems have tactics for automatic proving the propositional tautologies (see Appendix~\ref{appx_diff_table})

%//Fig.\ref{ex_morgan_propos_coq} shows the proof of De Morgan's law in Coq system. The law statement is being proved for all propositions of type \texttt{Prop}  <which is ...> ... . For arguments of type \texttt{bool} the proof is trivial.

The proof in Coq can be much simpler if the theorem is formulated with usage of \textit{Set}-type \texttt{bool} (see definition of \texttt{bool} in Appendix~\ref{apx:type_definitions} Fig.\ref{ex_typedef_bool_coq}, see proof in Appendix~\ref{apx:morgan_propos} Fig.\ref{ex_morgan_bool_coq}). There, it is possible to use the tactic \texttt{destruct} to decompose type to different goals and prove them separately (in Coq, the '\texttt{;}' operator between two tactics instructs interpreter to apply next tactic to all subgoals produced by previous tactic call). Note that when the theorem was formulated in terms of variables of meta-type \texttt{Set}, the automatic tactic \texttt{tauto} fails, as it works only with propositions of meta-type \texttt{Prop}.

%------------------------------------------------------------

\subsubsection{Proofs in first-order logic}

As an example of proof in first-order logic, the first-order quantified de Morgan's laws have been chosen. In both Coq and Isabelle, the proof necessarily relies on the axiom of excluded middle as the \textit{existence} of an element is to be proven\footnote{in contract to the previous proofs formulated in propositional logic, where the existence of both propositions was assumed.}. The proof in Isabelle is written as a \textit{backward proof} (see Appendix~\ref{apx:morgan_firstorder} Fig.\ref{ex_morgan_quant_isabelle}). The Coq's proof imports the library \texttt{Coq.Logic.Classical\_Prop}, which contains definitions of classical logic, which are useful to extend intuitionistic logic to classical logic (see Appendix~\ref{apx:morgan_firstorder} Fig.\ref{ex_morgan_quant_coq}).

%// see slide 53 from %https://www.labri.fr/perso/casteran/CoqArt/Tsinghua/C3.pdf
%// "Predicates : a Predicate is just any function of type A1→A2 . . .An→Prop where Ai : Set for each i. Predicates are declared as any other function symbol"

%// to prove de morgan form of predicats with first-order quantifiers, we need to include the classical axiom of excluded middle

%------------------------------------------------------------

\subsubsection{Proofs using inductive types}

In both Isabelle and Coq, the natural numbers type \texttt{nat} is defined inductively on induction on zero as in Peano arithmetic (see Appendix~\ref{apx:type_definitions} Fig.\ref{ex_nat_definition_isabelle} and Fig.\ref{ex_nat_definition_coq}). As an example of statement with the type \texttt{nat}, the simple formula 
$2 \cdot S_{n} = {n \cdot (n + 1)}$ for sum $S_{n}$ of first $n$ integer numbers has been chosen (see proof in Isabelle in Appendix~\ref{apx:sum_inductive} Fig.\ref{ex_nat_sum_isabelle}, see proof in Coq in Appendix~\ref{apx:sum_inductive} Fig.\ref{ex_nat_sum_coq}). Note that the proof in Coq uses the library \texttt{Coq.omega.Omega}, which contains powerful tactics to simplifying and proving natural numbers formulas.

\subsubsection{Code extraction in Coq}
Furthermore, after the correctness of defined function \texttt{range\_sum} has been proven, it is possible to extract from Coq the verified function code in Haskell or Ocaml:

\begin{raggedleft}
\begin{tabular}{p{.48\columnwidth} p{.48\columnwidth}}
\begin{lstlisting}[caption={Extracted function in Haskell}]
range_sum :: Nat -> Nat
range_sum n =
  case n of {
    O -> O;
    S p -> add (range_sum p) (S p)}
\end{lstlisting} % empty lines after code and [*$ $*] are needed for caption alignment
&
\begin{lstlisting}[caption={Extracted function in OCaml}]
(** val range_sum : nat -> nat **)

let rec range_sum = function
    | O -> O
    | S p -> add (range_sum p) (S p)
\end{lstlisting}
\end{tabular}
\end{raggedleft}

%extraction is mainly performing a straightforward syntactic translation

%------------------------------------------------------------

%// Sometimes Isabelle advices how it is possible to prove a statement more easily
%иногда Подсказывает, как можно проще доказать
%напр, для ассоциативности дизъюнкции, lemma disj_swap: "P ∨ Q ⟹ Q ∨ P"
%proof (prove)
%goal (1 subgoal):
%1. P ∨ Q ⟹ Q ∨ P 
%Auto solve_direct: the current goal can be solved directly with
%Meson.disj_comm: ?P ∨ ?Q ⟹ ?Q ∨ ?P
%// overview good at: https://pdfs.semanticscholar.org/95bf/a1bf0cbf5ae2c2a70daa13d4966143bd96f8.pdf

% see https://people.cs.kuleuven.be/~bart.jacobs/coq-essence.pdf, 11 Coq versus classical logic

\subsection{Results of comparison}
\label{sec:joint_comparison}

In this paper, the authors have made an attempt to compare to different theorem provers, Coq and Isabelle/HOL, and both of them have been found highly developed and valuable, although they both require deep understanding of metamathematical concepts of the proof process. The list below summarises the main features of these two tools that the authors have noticed.

\begin{itemize}
	\itemsep0em
  
	\item \textit{Expressiveness of underlying logic:}
    \begin{subitemize}
    	\item Isabelle/HOL uses classical higher-order logic;
      \item Coq uses intuitionistic logic based on Calculus of Inductive Constructions theory, but may be extended to classical logic by assuming the axiom of excluded middle.
    \end{subitemize}
  
  \item \textit{Necessary background for using the theorem prover:}
  \begin{subitemize}
    \item From the author's personal point of view, Coq requires deeper understanding of underlying logic theory, since usually the intuitionistic logic is being studying as a further development of classical logic that adds large number of additional constraints to it;
    \item Nonetheless, the whole proof process may seem unfamiliar for users with traditional mathematical background, so that for these users both systems require large amount of additional learning (at least, understanding and memorising the most common tactics is least necessary requirement for using these systems).
  \end{subitemize}

  \item \textit{The level of the proof automation:}
  \begin{subitemize}
    \item Both systems have automatic tactics for proving (e.g., \texttt{auto} in Isabelle; \texttt{auto}, \texttt{tauto} in Coq) or simplification complex statements (e.g., automatic reasoner \texttt{blast} in Isabelle; automatic tactics \texttt{simpl}, \texttt{omega} in Coq). However, in some cases these tactics offer insufficient level of automation, particularly in proving theorems over natural numbers (see example in Appendix~\ref{apx:sum_inductive} Fig.\ref{ex_nat_sum_coq}, where numerous steps for rewriting the equation by calling \texttt{rewrite} had been performed in order to apply automatic tactic \texttt{omega}).
  \end{subitemize}
  
  \item \textit{Size of proof:}
  \begin{subitemize}
    \item Analogous proofs have approximately equal size in both systems, caeteris paribus.
  \end{subitemize}

 \item \textit{Number of supporting theories:}
  \begin{subitemize}
    \item Both Isabelle and Coq have rather large set of libraries containing formalised theories and data structures, that are being constantly replenished, see~\cite{tool_Isabelle} and~\cite{tool_Coq}.
  \end{subitemize}
  
  \item \textit{Expressiveness of syntax:}
    \begin{subitemize}
      \item Both systems have the built-in powerful functional languages, which can be used to define complex recursive structures;
      \item Both systems accept forward proofs (written in imperative style as a sequence of tactics calls). This method may seem non-natural mathematically, as the search for proof is being performed "blindly", preserving the goal of the implicitly;
      
      \item In contrast to Coq, the backward proof supported by Isabelle firstly states the target goal explicitly for every tactic (with keywords \texttt{show}, \texttt{have}, \texttt{assume}, etc.), so that the proof become much more readable, yet it requires more time to be written.
    \end{subitemize}
  
  \item \textit{Usability of the syntax:}
    \begin{subitemize}
      \item Although Isabelle recognises common mathematical ASCII symbols in proof which makes it much more readable, it may seem inconvenient to use them within IDE (e.g., character \isabelleinline{\<forall>} is incoded as \texttt{\textbackslash<forall>}, \isabelleinline{\<Sum>} as \texttt{\textbackslash<Sum>}, etc.);
      \item The syntax of Coq is closer to the syntax of a programming language rather than mathematics, apparently it was designed for convenient work with a keyboard.
    \end{subitemize}
  
  \item \textit{Usability of the native IDE:}
  \begin{subitemize}
    \item The authours are inclined to consider the Isabelle's jEdit Prover IDE more user-friendly as the whole proof is being recompiled every time user changes the syntax tree, which facilitates user to acquire the proof state for any arbitrary step of the proof;
    \item In contrast, the CoqIDE can change the proof state backward and forward linearly, which however implies less system overload.
  \end{subitemize}

  \item \textit{Additional comparison information:}
  \begin{subitemize}
    \item Coq has an essential feature that distincts it from most other theorem provers: it can extract the verified code for which compliance with the specification have been proved in a constructive way. This encourages using Coq as a software verification tool.
  \end{subitemize}

\end{itemize}

%============================================================

\section{Future work}
\label{sec:future_work}

Although this paper does not pretend to give a fully exhaustive comparative analysis of two such complex systems as Coq and Isabelle, the authors hope it will help users without advanced background in mathematics to be involved into the work with proof assistants more quickly and easily. In future, this paper tends to be a foundation for more advanced survey of automatic tools used in software verification.

%============================================================

\section*{Acknowledgements}
\label{sec:ack}

I wish to thank Prof. Stavros Tripakis for letting me dive into the exciting world of Logic, for providing feedback on my paper at all stages of the work, for answering all my countless questions and supporting me.

%============================================================

\bibliographystyle{ieeetr}
\bibliography{cs-seminar}

\begin{thebibliography}{10}

\bibitem{tool_Isabelle}
``Isabelle, a generic proof assistant.''
\newblock \url{https://www.cl.cam.ac.uk/research/hvg/Isabelle/}.

\bibitem{tool_Coq}
``Coq proof assistant.''
\newblock \url{https://coq.inria.fr/}.

\bibitem{tool_Pvs}
``{PVS} specification and verification system.''
\newblock \url{http://pvs.csl.sri.com/}.

\bibitem{tool_Otter}
W.~McCune, ``{Otter} and {Mace2},'' 2003.
\newblock \url{https://www.cs.unm.edu/~mccune/otter/}.

\bibitem{tool_Acl}
``{ACL2}: a computational logic for applicative common lisp.''
\newblock \url{http://www.cs.utexas.edu/users/moore/acl2/}.

\bibitem{Wie03}
F.~Wiedijk, ``Comparing mathematical provers,'' in {\em MKM}, vol.~3,
  pp.~188--202, Springer, 2003.

\bibitem{Griff98}
D.~Griffioen and M.~Huisman, ``A comparison of pvs and isabelle/hol,'' {\em
  Theorem Proving in Higher Order Logics}, pp.~123--142, 1998.

\bibitem{Roj15}
R.~Rojas, ``A tutorial introduction to the lambda calculus,'' {\em arXiv
  preprint arXiv:1503.09060}, 2015.

\bibitem{Bar88}
H.~P. Barendregt, ``Introduction to lambda calculus,'' 1988.

\bibitem{Bar13}
H.~Barendregt, W.~Dekkers, and R.~Statman, {\em Lambda calculus with types}.
\newblock Cambridge University Press, 2013.

\bibitem{Tho91}
S.~Thompson, {\em Type theory and functional programming}.
\newblock Addison Wesley, 1991.

\bibitem{Irv95}
A.~D. Irvine and H.~Deutsch, ``Russell's paradox,'' {\em Stanford encyclopedia
  of philosophy}, 2008.

\bibitem{Paulson90}
L.~C. Paulson, ``A formulation of the simple theory of types (for isabelle),''
  in {\em COLOG-88}, pp.~246--274, Springer, 1990.

\bibitem{Rice53}
H.~G. Rice, ``Classes of recursively enumerable sets and their decision
  problems,'' {\em Transactions of the American Mathematical Society}, vol.~74,
  no.~2, pp.~358--366, 1953.

\bibitem{Coq85}
T.~Coquand, {\em Une Théorie des Constructions}.
\newblock PhD thesis, Université de Paris VII, 1995.

\bibitem{Paulin15}
C.~Paulin-Mohring, ``Introduction to the calculus of inductive constructions,''
  2015.

\bibitem{Pierce2002}
B.~C. Pierce, {\em Types and programming languages}.
\newblock MIT press, 2002.

\bibitem{Wiedijk100}
F.~Wiedijk, ``Formalizing 100 theorems.''
\newblock \url{http://www.cs.ru.nl/~freek/100/}.

\bibitem{tool_HOL}
``{HOL}, interactive theorem prover.''
\newblock \url{https://hol-theorem-prover.org/}.

\bibitem{Letouzey08}
P.~Letouzey, ``Extraction in {Coq}: An overview,'' {\em Logic and Theory of
  Algorithms}, pp.~359--369, 2008.

\bibitem{Klein09}
G.~Klein, K.~Elphinstone, G.~Heiser, J.~Andronick, D.~Cock, P.~Derrin,
  D.~Elkaduwe, K.~Engelhardt, R.~Kolanski, M.~Norrish, T.~Sewell, H.~Tuch, and
  S.~Winwood, ``sel4: Formal verification of an os kernel,'' in {\em
  Proceedings of the ACM SIGOPS 22Nd Symposium on Operating Systems
  Principles}, SOSP '09, (New York, NY, USA), pp.~207--220, ACM, 2009.

\bibitem{Krebbers15}
R.~J. Krebbers, {\em The C standard formalized in Coq}.
\newblock [Sl: sn], 2015.

\bibitem{tool_PG}
``Proof general.''
\newblock \url{https://proofgeneral.github.io/}.

\bibitem{Glivenko29}
V.~Glivenko, ``Sur quelques points de la logique de m. brouwer,'' {\em
  Bulletins de la classe des sciences}, vol.~15, no.~5, pp.~183--188, 1929.

\bibitem{Kolmogorov25}
A.~Kolmogorov, ``O principe tertium non datur (in russian),'' {\em English
  trans. in van Heijenoort [1967, 414-437], from Mathematicheskiy sbornik 32:
  646--667}, 1925.

\bibitem{Rush14}
P.~Rush, {\em The Metaphysics of Logic}.
\newblock Cambridge University Press, 2014.

\bibitem{Harrison09}
J.~Harrison, {\em Handbook of practical logic and automated reasoning}.
\newblock Cambridge University Press, 2009.

\end{thebibliography}

%============================================================

\newpage
\appendix
%\addcontentsline{toc}{section}{Appendices}
\section*{Appendices}
\addcontentsline{toc}{section}{Appendices}
\renewcommand{\thesubsection}{A.\arabic{subsection}}

\subsection{Basic type definitions}
\label{apx:type_definitions}

\begin{lstlisting}[language=coq,
  caption={Basic \texttt{Prop} types definitions in Coq},
  label={ex_typedef_prop_coq}]
(* In Coq, False is an unobservable proposition, which
   is defined as a propositional type without constructor *)
Inductive False : Prop := .

(* On the other hand, True is defined as always true proposition *)
Inductive True : Prop := I : True.
\end{lstlisting}

\begin{lstlisting}[language=coq, 
  caption={Basic \texttt{Set} types definitions in Coq},
  label={ex_typedef_bool_coq}]
(* boolean type is defined as simple enumeration: *)
Inductive bool : Set :=
    true  : bool | false : bool

(* Similartly to the False, an empty set is a Set without type constructor: *)
Inductive Empty_set : Set := .
\end{lstlisting}

\begin{raggedleft}
\begin{tabular}{p{.49\linewidth} p{.48\linewidth}}
\begin{lstlisting}[language=isabelle, caption={Definition of Peano's natural numbers type \texttt{nat} in~Isabelle}, label={ex_nat_definition_isabelle}]
datatype nat = 
    zero ("0") |
    Suc nat
\end{lstlisting}
&
\begin{lstlisting}[language=coq, caption={Definition of Peano's natural numbers type \texttt{nat} in~Coq}, label={ex_nat_definition_coq}]
Inductive nat : Type :=
    | O : nat
    | S : nat -> nat.
\end{lstlisting}
\end{tabular}
\end{raggedleft}

\begin{raggedleft}
  \begin{tabular}{p{.49\linewidth} p{.49\linewidth}}
    \begin{lstlisting}[language=isabelle, caption={Definition of addition over \texttt{nat} in~Isabelle}]
fun add :: "nat => nat => nat"
  where
    "add 0 n = n" |
    "add (Suc m) n = Suc(add m n)"

[* $ $ *]
    \end{lstlisting}
    &
    \begin{lstlisting}[language=coq, caption={Definition of addition over \texttt{nat} in~Coq}]
Fixpoint add (n m: nat) : nat :=
    match n with
        | O => m
        | S n' => S (n' + m)
    end
where "n + m" := (add n m) : nat_scope.
    \end{lstlisting}
  \end{tabular}
\end{raggedleft}

\newpage
\subsection{Example proof of double-negated classical tautology in Coq}
\label{apx:double_negated}

%\begin{minipage}[t]{\linewidth}
%, float,floatplacement=H
\begin{lstlisting}[language=coq,
    caption={Proof of the double-negated excluded middle in Coq},
    label={ex_double_neg_ex_mid_coq}]
Lemma DoubleNegatedExcludedMiddle_Coq: forall P: Prop,
    ~~(P \/ ~P).
Proof.
    (* 'tauto' automatically proves the statement *)
    unfold not.    (* apply ~P ==> P -> False *)
    intros P f.    (* move premises to the set of hypotheses *)
    apply f.       (* replace the goal with premise of implication in f *)
    right.         (* apply disjunction elimination inference rule *)
    intro P_holds. (* move P to the set of hypotheses *)
    apply f.       (* replace the goal with premise of implication in f *)
    left.          (* apply disjunction elimination inference rule *)
    exact P_holds. (* match the goal with one of the hypotheses *)
Qed.
\end{lstlisting}
%\end{minipage}

\newpage
\subsection{Example proofs of de Morgan's laws in propositional logics}
\label{apx:morgan_propos}

%In proofs, some steps are followed by the comments (holding between symbols '\texttt{(*}' and '\texttt{*)}') that show how the tactic has changed the proof state.
%//For Isabelle, same de Morgan's law for propositions:
% source: https://www.inf.ed.ac.uk/teaching/courses/a	r/isabelle/exercises/propositional/sol.pdf
\begin{lstlisting}[language=isabelle, caption={Proof of the de Morgan's law for propositions in Isabelle}, label={ex_morgan_propos_isabelle}]
lemma DeMorganPropositional_Isabelle:
  "(\<not> (P \<and> Q)) = (\<not> P \<or> \<not> Q)"
  
  (* 'apply blast' automatically solves the equation *)
  apply (rule iffI)      (* split equality into two subgoals *)
  (* "Forward" subgoal: 1. \<not>(P \<and> Q) ==> \<not> P \<or> \<not> Q *)
  apply (rule classical) (* 1. \<not> (P \<and> Q) ==> \<not> (\<not> P \<or> \<not> Q) ==> \<not> P \<or> \<not> Q *)
  apply (erule notE)     (* 1. \<not> (\<not> P \<or> \<not> Q) ==> P \<and> Q *)
  apply (rule conjI)     (* 1. \<not> (\<not> P \<or> \<not> Q) ==> P; 2. \<not> (\<not> P \<or> \<not> Q) ==> Q *)
  apply (rule classical) (* 1. \<not> (\<not> P \<or> \<not> Q) ==> \<not> P ==> P *)
  apply (erule notE)     (* 1. \<not> P ==> \<not> P \<or> \<not> Q *)
  apply (rule disjI1)    (* 1. \<not> P ==> \<not> P *)
  apply assumption       (* 1. (solved). 2. \<not> (\<not> P \<or> \<not> Q) ==> Q *)
  apply (rule classical) (* 2. \<not> (\<not> P \<or> \<not> Q) ==> \<not> Q ==> Q *)
  apply (erule notE)     (* 2. \<not> Q ==> \<not> P \<or> \<not> Q *)
  apply (rule disjI2)    (* 2. \<not> Q ==> \<not> Q *)
  apply assumption       (* 2. (solved) *)
  (* "Backward" subgoal: 3. \<not> P \<or> \<not> Q ==> \<not> (P \<and> Q) *)
  apply (rule notI)   (* 3. \<not> P \<or> \<not> Q ==> P \<and> Q ==> False *)   
  apply (erule conjE) (* 3. \<not> P \<or> \<not> Q ==> P ==> Q ==> False *)
  apply (erule disjE) (* 3. P ==> Q ==> \<not>P ==>False; 4. P ==> Q==> \<not>Q==> False *)
  apply (erule notE, assumption)+  (* 3. (solved); 4. (solved) *)
done
\end{lstlisting}

\begin{lstlisting}[language=coq, caption={Proof of the de Morgan's law for propositions in Coq}, label={ex_morgan_propos_coq}]
Theorem DeMorganPropositional_Coq:
    forall P Q : Prop, ~(P \/ Q) <-> ~P /\ ~Q.  
Proof.
  (* 'tauto' automatically proves the equation *)
  intros P Q. unfold iff.
  split.
  - intros H_not_or. unfold not. constructor.
    + intro H_P. apply H_not_or. left. apply H_P.
    + intro H_Q. apply H_not_or. right. apply H_Q.
  - intros H_and_not H_or.
    destruct H_and_not as [H_not_P H_not_Q].
    destruct H_or as [H_P | H_Q].
    + apply H_not_P. assumption.
    + apply H_not_Q. assumption.
Qed.
\end{lstlisting}

%<here bool the datatype, which is inductively defined as a \texttt{Set} => can use destruct, which expands the definition of inductive type>
\begin{lstlisting}[language=coq,caption={Proof of the de Morgan's law for booleans in Coq}, label={ex_morgan_bool_coq}]
(* define macroses: *)
Notation "a || b" := (orb a b).
Notation "a && b" := (andb a b).
Theorem DeMorganBoolean_Coq:
    forall a b: bool, negb (a || b) = ((negb a) && (negb b)).
Proof.
    try tauto.  (* automatic tactic fails here *)
    intros a b.
    destruct a; simpl; reflexivity.
Qed.
\end{lstlisting}

\newpage
\subsection{Example proofs of first-order quantified de Morgan's laws}
\label{apx:morgan_firstorder}

%https://www.cl.cam.ac.uk/research/hvg/Isabelle/dist/Isabelle/browser_info/HOL/HOL-Proofs-ex/Drinker.html
\begin{lstlisting}[language=isabelle, caption={Proof of the de Morgan's law for first-order propositions in Isabelle}, label={ex_morgan_quant_isabelle}]
lemma DeMorganQuantified_Isabelle[*\footnote{This proof was originally taken from the set of examples in Isabelle's documentation, see \\ \footnotesize{https://github.com/seL4/isabelle/blob/master/src/HOL/Isar\_Examples/Drinker.thy}}*]:
  assumes "\<not> (\<forall>x. P x)"
  shows "\<exists>x. \<not> P x"
proof (rule classical)
  assume "\<nexists>x. \<not> P x"
  have "\<forall>x. P x"
  proof
    fix x show "P x"
    proof (rule classical)
      assume "\<not> P x"
      then have "\<exists>x. \<not> P x" ..
      with <\<nexists>x. \<not> P x> show ?thesis by contradiction
    qed
  qed
  with <\<not>(\<forall>x. P x)> show ?thesis by contradiction
qed
\end{lstlisting}

% HONESTLY I TOOK THE PROOF FROM http://flint.cs.yale.edu/cs428/coq/library/Coq.Logic.Classical_Pred_Type.html
\begin{lstlisting}[language=coq, caption={Proof of the de Morgan's law for first-order propositions in Coq}, label={ex_morgan_quant_coq}]
Require Import Coq.Logic.Classical_Prop.

Lemma DeMorganQuantified_Coq: forall (P : Type -> Prop), 
    ~ (forall x : Type, P x) -> exists x : Type, ~ P x.
Proof.
    unfold not.
    intros P H_notall.
    apply NNPP.  (* apply classic rule ~~P ==> P *)
    unfold not. intro H_not_notexist.
    cut (forall x:Type, P x).  (* add new goal from the goal's premise *)
    - exact H_notall.
    - intro x. apply NNPP.
      unfold not.
      intros H_not_P_x.
      apply H_not_notexist.
      exists x.
      exact H_not_P_x.
Qed.
\end{lstlisting}

\newpage
\subsection{Example of higher-order statement definitions}

\begin{raggedleft}
  \begin{tabular}{p{.48\linewidth} p{.48\linewidth}}
    \begin{lstlisting}[language=isabelle, caption={Higher-order statement definition in Isabelle}] 
lemma lem:
  "\<forall> (f::bool=>bool) (b::bool) .
  f (f (f b)) = f b"
    \end{lstlisting}
    &
    \begin{lstlisting}[language=coq, caption={Higher-order statement definition in Coq}]
Lemma lem: 
    forall (f : bool -> bool) (b : bool),
    f (f (f b)) = f b.
    \end{lstlisting}
  \end{tabular}
\end{raggedleft}

\newpage
\subsection{Example proofs of the formula for sum of first \textit{n} numbers using inductive types}
\label{apx:sum_inductive}

\begin{lstlisting}[language=isabelle, caption={Proof of the formula for sum of n first number in Isabelle}, label={ex_nat_sum_isabelle}]
fun range_sum :: "nat => nat"
  where "range_sum n = (\<Sum>k::nat=0..n . k)"
value "range_sum 10"  (* check the function *)

theorem SimpleArithProgressionSumFormula_Isabelle: "2 * (range_sum n) = n * (n + 1)"
  proof (induct n)
    show "2 * range_sum 0 = 0 * (0 + 1)" by simp
  next
  fix n have "2 * range_sum (n + 1) = 2 * (range_sum n) + 2 * (n + 1)" by simp
  also assume "2 * (range_sum n) = n * (n + 1)"
  also have "\<dots> + 2 * (n + 1) = (n + 1) * (n + 2)" by simp
  finally show "2 * (range_sum (Suc n)) = (Suc n) * (Suc n + 1)" by simp
qed
\end{lstlisting}

\begin{lstlisting}[language=coq, caption={Proof of the formula for sum of n first number in Coq}, label={ex_nat_sum_coq}]
Require Import Coq.[*omega*].Omega.
Require Coq.Logic.Classical.

Fixpoint range_sum (n: nat) : nat :=
    match n with
        | O => 0
        | S p => range_sum p + (S p)
    end.
Compute range_sum 3.  (* output: '= 6 : nat' *)

Lemma range_sum_lemma: forall n: nat,
    range_sum (n + 1) = range_sum n + (n + 1).
Proof.
    intros. induction n.
    - simpl; reflexivity.
    - simpl; omega.
Qed.

Theorem SimpleArithProgressionSumFormula_Coq:
    forall n, 2 * range_sum n = n * (n + 1).
Proof.
    intros.
    induction n.
    (* goal: '2 * range_sum 0 = 0 * (0 + 1)' *)
    - simpl; reflexivity.
    (* goal: '2 * range_sum (S n) = S n * (S n + 1)' *)
    - rewrite -> Nat.mul_add_distr_l. (* '2*range_sum(S n) = S n * S n + S n * 1' *)
      rewrite -> Nat.mul_1_r.         (* '2*range_sum(S n) = S n * S n + S n' *)
      rewrite -> (Nat.mul_succ_l n).  (* '2*range_sum(S n) = n * S n + S n + S n' *)
      rewrite <- (Nat.add_1_r n).     (* '2*range_sum(n+1) = n*(n+1)+(n+1)+(n+1)' *)
      rewrite -> range_sum_lemma.     (* '2*(range_sum(n)+(n+1)) = n*(n+1)+(n+1)+(n+1)' *)
      omega.                          (* automatically solve arithmetic equation *)
Qed.
\end{lstlisting}

%// TODO: Inductive datatype
%Isabelle:"datatype ’a list = Nil | Cons ’a (’a list)" %https://www.it.uu.se/education/phd_studies/phd_courses/gc0910/isabelle/slide2.pdf

\end{document}